\documentclass[amssymb,
               amsmath,
               aps,
               preprint,
               preprintnumbers,
               showpacs,
               showkeys
              ]{revtex4}

\usepackage{bm}
\usepackage{graphicx}
\usepackage{epsfig}

\def\lowmp{\lower.11em\hbox{${\scriptstyle\mp}$}}

\def\>{\mskip\medmuskip}
\def\E3{\mathbb{E}^3}
\def\R3{\mathbb{R}^3}
\def\T3{\mathbb{T}^3}
\def\V3{\mathbb{V}^3}

\newcommand{\vt}[1]{\mathbf{#1}}

\begin{document}

\title{A simple approach to the calculation of retarded dispersion forces}
\date{\today}
\author{B Carazza$^1$ and R Tedeschi$^2$}
\address{$^1$Department of Physics University of Parma, \\ 
Parco Area delle Scienze 7/A I43100 Parma, Italy        \\
I.N.F.N. Sezione di Cagliari, Italy                     \\
$^2$Department of Physics University of Parma,          \\ 
Parco Area delle Scienze 7/A I43100 Parma, Italy        \\
I.N.F.N. Gruppo Collegato di Parma, Italy               \\
carazza@fis.unipr.it, tedeschi@fis.unipr.it}
\begin{abstract}
We propose a phenomenological Hamiltonian for the interaction of
neutral macroscopic bodies with the electromagnetic field. Subsequently
we revisit the assumption according to which the retarded interactions
between neutral macroscopic bodies can be obtained through an additive
principle i.e. summing the volume elements defined by the macroscopic 
bodies.
\end{abstract}



\maketitle

\date{\today}

\section{Introduction}
As it is very well known the discovery and the first calculation of the 
retarded forces between macroscopic bodies, due to the fluctuations of 
the electromagnetic field, can be found in Casimir's 1948 seminal 
paper\cite{Casimir1}. A few years later the Lifshitz theory appeared.
Within the frame of this theory a powerful expression describing the force
at finite temperature between two semiinfinite dispersive media separated 
by a slab was obtained\cite{Lifshitz}. However the fundamental method of 
calculating these interactions at zero temperature is still due 
to Casimir. One has to single-out the electromagnetic field normal 
modes consistent with the boundary conditions imposed by the presence 
of the macroscopic bodies. The shift of the zero point energy of the 
field with respect to vacuum is then considered as the sought interaction 
energy. In the case of complicated geometry the evaluation of the Casimir 
force is very difficult, therefore various approximation methods were 
developed such as the so called proximity-force approximation 
(PFA)\cite{Derjaguin}, the semiclassical\cite{Schaden,Mazzittelli} 
and optical approximations\cite{Jaffe} and path-integral methods\cite{Gies}.
We are interested here in an approach used in reference \cite{Abrikosova} 
and discussed by Mostepanenko and Sokolov\cite{Mostepanenko}. The 
interested reader can find a review of the subject in the comprehensive
report by Bordag et al.\cite{Bordag}. It is assumed that the dependence 
of the interaction potential of macroscopic bodies on distance can 
be determined by summing the interaction potential between single atoms 
(molecules) belonging to different bodies. The coefficient of the 
expression so obtained, which depends on the kind of material involved, 
is renormalized in order to be in agreement with the interaction
between flat plates of the same material as derived by the Lifshitz theory.
The afore mentioned additivity principle works very well in the case of
low density media though it is not said that the principle could be
extended in general and particularly in the case of metals. However
because of the simplicity and usefullness of this approach we are here
revisiting it with the idea according to which the retarded forces between 
two macroscopic bodies might be understood as being due to the attraction 
in pairs between their volume elements. Evidently such volume elements must 
be considered small as far as the differential calculus is concerned, yet 
rather large compared to interatomic distances. We will delineate our idea 
proposing a phenomenological Hamiltonian for the interaction of the 
electromagnetic field with a macroscopic body whose properties are 
phenomenologically described. 
 
In the following sections we will discuss the proposed Hamiltonian
showing where it leads. Subsequently we will apply our
results to a few geometrical dispositions of simple macroscopic objects.

\section{A phenomenologica perturbation Hamiltonian}
Consider two neutral, dielectric and homogeneous bodies in fixed positions.
The presence of such bodies will modify the Hamiltonian of the 
electromagnetic field which will be written:
\begin{equation}
   {\cal{H}}={\cal{H}}_0+{\cal{H}'}
\label{10}
\end{equation}
where ${\cal{H}}_0$ is the Hamiltonian of the free electromagnetic field
and where ${\cal{H}'}$ indicates the perturbation due to the presence of the
two bodies. Assuming as unperturbed field state the vacuum we will calculate
the energy shift due to the extra term of the Hamiltonian. The part of the 
shift which depends on the configuration of the two bodies will be interpreted
as their potential energy at zero temperature. We consider and propose now
a phenomenological perturbation as follows:
\begin{equation}
   {\cal{H}'}=-\frac{1}{2}\int_V\vt{P}\cdot\vt{E}\,dV
              -\frac{1}{2}\int_V\vt{M}\cdot\vt{H}\,dV
\label{20}
\end{equation}
where $\vt{E}$, $\vt{H}$ are the electric and magnetic fields, $\vt{P}$ and
$\vt{M}$ are the polarization and magnetization vectors.
The integral is extended to the volume occupied by the bodies. 
Always within a phenomenological approach the Fourier components 
$\vt{P}_{\vt{k}}$ and $\vt{M}_{\vt{k}}$ may be considered connected
to those of the electric field and magnetic field by a linear relation.

We set:
\begin{equation}
   \begin{cases}
     \hat{\vt{P}}=\displaystyle{i\sqrt{2\pi\hslash c}
                  \sum_{\vt{k},\lambda}\sqrt{k}\beta(k)
                  \left( a_{\vt{k},\lambda}e^{i\vt{k}\cdot\vt{r}}-
                         a^\dagger_{\vt{k},\lambda}e^{-i\vt{k}\cdot\vt{r}}
                  \right)\vt{e}_\lambda(\vt{k})} \\
     \hat{\vt{M}}=\displaystyle{i\sqrt{2\pi\hslash c}
                  \sum_{\vt{k},\lambda}\sqrt{k}\gamma(k)
                  \left( a_{\vt{k},\lambda}e^{i\vt{k}\cdot\vt{r}}-
                         a^\dagger_{\vt{k},\lambda}e^{-i\vt{k}\cdot\vt{r}}
                  \right)
                  \frac{\vt{k}\times\vt{e}_\lambda(\vt{k})}{k}}
   \end{cases}
\label{30}
\end{equation}
where $a_{\vt{k}}$, $a^\dagger_{\vt{k}}$ are the annihilation and creation 
operators and $\vt{e}_\lambda(\vt{k})$ indicates the polarization. For the
field we have used only transversal waves and we set to unity the side of
the cube which defines the periodic conditions. The $\beta(k)$ and 
$\gamma(k)$ coefficients are defined as:
\begin{equation}
   \begin{cases}
   \beta(k)=\displaystyle{
            \frac{3}{4\pi}\frac{\varepsilon(k)-1}{\varepsilon(k)+2}}\\
   \gamma(k)=\displaystyle{
             \frac{3}{4\pi}\frac{\mu(k)-1}{\mu(k)+2}}
\end{cases}
\label{50}
\end{equation}
where $\varepsilon$ and $\mu$ are the electric and magnetic permittivities
which expresses the usual Clausius-Mossotti formula. We then obtain:
\begin{multline}
   {\cal{H}'}=\pi\hslash c
              \sum_{\vt{k}\cdot\vt{k}'}
              \sum_{\lambda,\nu}
              \sum_i\sqrt{kk'}
              \left[\vt{e}_\lambda(\vt{k})\cdot\vt{e}_\nu(\vt{k}')\beta(k)+
                    \frac{\vt{k}\times\vt{e}_\lambda(\vt{k})}{k}\cdot
                    \frac{\vt{k'}\times\vt{e}_\nu(\vt{k'})}{k'}
                    \gamma(k)\right] \\
     \left[a_{\vt{k},\lambda}a_{\vt{k}',\nu}v_i(\vt{k}+\vt{k}')-
      a^\dagger_{\vt{k},\lambda}a_{\vt{k}'\nu}v_i(\vt{k}'-\vt{k})-
      a_{\vt{k},\lambda}a^\dagger_{\vt{k}'\nu}v_i(\vt{k}-\vt{k}')+
      a^\dagger_{\vt{k},\lambda}a^\dagger_{\vt{k}'\nu}v_i(-\vt{k}-\vt{k}')
     \right]
\label{60}
\end{multline}
The index $i=1,2$ indicates the summation over the two bodies. We 
furthermore defined:
\begin{equation}
   v_i(\bm{\eta})=\int_{V_i}e^{i\bm{\eta}\cdot\vt{r}_i}dV_i
\label{70}
\end{equation}

\section{The interaction energy at large separations}
We stress that we consider the perturbation $\cal{H}'$ above to hold only 
for the calculation of the interaction energy of neutral macroscopic bodies 
at large separations. Let $d$ be the shortest distance between the surfaces 
of the two bodies. The expression ``large separation'' usually means 
$d >> \lambda_i$ where $\lambda_i$ are the absorption wavelengths of 
the material under consideration.
The mutual energy of the two bodies at zero temperature is obtained by
perturbation at first order. In the case of large separations the 
calculation simplifies greatly because $\beta(k)$ and $\gamma(k)$ 
can be safely substituted by $\beta(0)$ and $\gamma(0)$. In fact the major 
contribution to the integral below comes from the $k\le 1/d$ region where 
$d$, as previously mentioned, is the shortest distance between the surfaces 
of the bodies\cite{Power}. 

Thus we asymptotically obtain:
\begin{multline}
   U_{1,2}=-\frac{\hslash c}{2(2\pi)^4}
      \left[\beta_1(0)\beta_2(0)+\gamma_1(0)\gamma_2(0)\right]\times\\
      \int d\vt{k}\int d\vt{k}'
      \frac{kk'}{k+k'}
      \left(1+\frac{(\vt{k}\cdot\vt{k}')^2}{k^2{k'}^2}\right)
      \left(v_1(\vt{k}+\vt{k}')v_2(-\vt{k}+\vt{k}')+c.c.\right)
\label{90}
\end{multline}

Setting $\vt{q}=\vt{k}+\vt{k}'$ the expression \eqref{90} becomes:
\begin{multline}
   U_{1,2}=-\frac{\hslash c}{2(2\pi)^4}
      \left[\beta_1(0)\beta_2(0)+\gamma_1(0)\gamma_2(0)\right]\times\\
      \int d\vt{q}(v_1(\vt{q})v_2(\vt{-q})+c.c.)
      \int d\vt{k}\frac{k|\vt{q}-\vt{k}|}{k+|\vt{q}-\vt{k}|}
      \left(1+\frac{(\vt{k}\cdot(\vt{q}-\vt{k}))^2}
                   {k^2|\vt{q}-\vt{k}|^2}\right)
\label{100}
\end{multline}
The integral over $\vt{k}$ because of a lack of a quenching factor 
obviously diverges. Yet within the scope of our calculation we can 
neglect the infinities and consider only the singular part in the 
origin $(q=0)$ which dominates the asimptotic behaviour. 
The present procedure is analogous to the usual renormalization techniques
which make use of a cut-off. In our case the divergent part of the integral
is already separated without ambiguities. Furthermore this part can be 
ignored because it does not depend on the system configuration and 
therefore it can not give origin to forces among the considered bodies.
The final result,indicated in expression \eqref{110}, can be obtained 
rather laboriously following the indications of reference \cite{Landau2}:
\begin{equation}
   U_{1,2}=\hslash c A_{1,2}\int d\vt{q}(v_1(\vt{q})v_2(\vt{-q})+c.c.)q^4\ln q
\label{110}
\end{equation}
where:
\begin{equation}
   A_{1,2}=\frac{23}{240(2\pi)^3}
           \left[\beta_1(0)\beta_2(0)+\gamma_1(0)\gamma_2(0)\right]
\label{115}
\end{equation}
For two perfect metals $[\varepsilon(0)=\infty$, $\mu(0)=0]$ the constant 
$A_{m,m}$ is:
\begin{equation}
   A_{m,m}=\frac{1035}{3840}\frac{1}{(2\pi)^5}
\label{117}
\end{equation}
Now using the tables of the Fourier transforms of the generalized functions 
\cite{Gel} we have:
\begin{equation}
   \int d\vt{q}q^4\ln qe^{i\vt{k}\cdot\vt{r}}=
   -\frac{30(2\pi)^3}{\pi r^7}
\label{120}
\end{equation} 
hence recalling \eqref{70} we obtain:
\begin{equation}
   U_{1,2}=-\hslash c B_{1,2}
      \int_{V_1} d\vt{r}_1\int_{V_2} d\vt{r}_2\frac{1}{|\vt{r}_1-\vt{r}_2|^7}
\label{130}
\end{equation}
where:
\begin{equation}
   B_{1,2}=\frac{23}{4\pi}\left[\beta_1(0)\beta_2(0)+
                                \gamma_1(0)\gamma_2(0)\right]
\label{135}
\end{equation}
In the case of two perfect metals the constant is:
\begin{equation}
   B_{m,m}=\frac{1035}{256}\frac{1}{\pi^3}
\label{137}
\end{equation}
As it can be seen from expression \eqref{130} our results confirm 
Mostepanenko's assumptions\cite{Mostepanenko} yielding 
in addition the coefficient as a function of the characteristics 
of the material. 

\section{A few examples}
In the case of two alike homogeneous dielectrics of very low molecular density 
and where the molecules are assumed of the same kind $\gamma(0)=0$ and
$\beta(0)$ becomes $N\alpha(0)$ where $N$ is the molecular density and 
$\alpha(0)$ the molecular polarization. From \eqref{130} we then have:
\begin{equation}
     U=-\frac{23}{4\pi}\hslash c\alpha_1(0)\alpha_2(0)
      \int N d\vt{r}_1\int N d\vt{r}_2
      \frac{1}{|\vt{r}_1-\vt{r}_2|^7} 
\label{160}
\end{equation}
Which means that the interaction potential of the retarded Van der Waals 
forces between two molecules is:
\begin{equation}
   -\frac{23}{4\pi}\hslash c\alpha_1(0)\alpha_2(0)
    \frac{1}{|\vt{r}_1-\vt{r}_2|^7} 
\label{170}
\end{equation}
which is a well known result.\cite{Power,Casimir2}

Shall we move on now to the classic case of two parallelepipeds made of 
perfect metal whose volumes are so defined:
\begin{alignat}{3}
   -\frac{L}{2} & \le x_1     \le \frac{L}{2}\qquad
   -\frac{L}{2} & \le y_1     \le \frac{L}{2}\qquad
             -D & \le z_1 \le 0                 \\ 
   -\frac{L}{2} & \le x_2     \le \frac{L}{2}\qquad
   -\frac{L}{2} & \le y_2     \le \frac{L}{2}\qquad
              a & \le z_2 \le a+D                
\label{190}
\end{alignat}
The volume Fourier transforms are:
\begin{equation}
   v_1(\vt{q})v_2(-\vt{q})=16
   \left(\frac{\sin\frac{q_xL}{2}}{q_x}\right)^2\times
   \left(\frac{\sin\frac{q_yL}{2}}{q_y}\right)^2\times 
         \frac{\left(1-e^{-iq_zD}\right)
               \left(e^{-iq_z(D+a)}-e^{-iq_za}\right)}{q_z^2}
\label{200}
\end{equation}
To integrate \eqref{200} we find it convenient, not to mix up the calculations,
to write:
\begin{equation}
   \left(\frac{\sin\frac{q_xL}{2}}{q_x}\right)^2=
   \frac{\sin\frac{q_xL}{2}}{q_x}\frac{\sin\frac{q_xL'}{2}}{q_x}
\label{210}
\end{equation}
regarding the first of the factors on the right (assuming $L$ large) as 
\begin{equation}
   \frac{\sin\frac{q_xL}{2}}{q_x}=\pi\delta(q_x)
\label{220}
\end{equation}
we then calculate the limit $L'\rightarrow L$ once the integration has been
carried out. We follow the same route with the factor depending on $q_y$. 
We are then led to the integral:
\begin{equation}
   (2\pi)^2L^2\int dq_zq^2_z\ln(q_z)
   \left[\left(2e^{-iq_z(a+D)}-e^{-iq_z(a+2D)}-e^{-iq_za}\right)+c.c.\right]
\label{230}
\end{equation}
which can be evaluated using again reference \cite{Gel}. One then finds:
\begin{equation}
   -2(2\pi)^3L^2\left(\frac{1}{a^3}+\frac{1}{(a+D)^3}-\frac{2}{(a+D)^3}\right)
\label{240}
\end{equation}
Inserting the previous result into expression \eqref{110} we obtain the
interaction energy for two metallic parallelepipeds at zero temperature:
\begin{equation}
   -\frac{1035}{7680}\frac{\hslash c}{\pi^2}L^2
    \left(\frac{1}{a^3}+\frac{1}{(a+2D)^3}-\frac{2}{(a+D)^3}\right)
\label{250}
\end{equation}
which at the limit $D\gg a$ becomes:
\begin{equation}
   -\frac{1035}{7680}\frac{\hslash cL^2}{a^3}=
   -0.01365\frac{\hslash cL^2}{a^3}
\label{260}
\end{equation}
Expression \eqref{260} has to be compared with Casimir's more elegant result
\cite{Casimir1}:
\begin{equation}
   -\frac{\pi^2}{720}\frac{\hslash cL^2}{a^3}=
   -0.01370\frac{\hslash cL^3}{a^3}
\label{270}
\end{equation}
The small discrepancy between the two results deserves a brief comment. 
Though  its origin is not clearly evident, we think that it comes from 
the fact that our Hamiltonian, referring to the bulk of the surrounding
bodies, does not produce an exact description of the field in the proximity of
the surface where the Clausiu-Mossotti formula may be inadeguate. In the
Casimir's approach the appropriate boundary conditions are generally 
referred tho the field, thus the description of the situation on the 
surface is certaninly more exact.

Consider now the interaction $u$ of a metallic corpuscle
of infinitesimal volume $dV$ at distance $y$ from a metallic hemispace. 
From expression \eqref{130} with $B_{1,2}=B_{m,m}$ and using cylindrical 
coordinates we have:
\begin{equation}
   u=-\hslash cB_{m,m}dV\int_0^{2\pi}d\varphi
            \int_0^\infty\rho d\rho
            \int_0^\infty dz
            \frac{1}{\left[\rho^2+(z+y)^2\right]^{-7/2}}
\label{290}
\end{equation}
which is:
\begin{equation}
   u=-\hslash cB_{m,m}\frac{\pi}{10}\frac{1}{y^4}dV
\label{300}
\end{equation}
Using this last expression one can calculate the interaction between a very
thick and infinitely extended metallic plate and a generic body. Let this
last be a metallic sphere of radius $R$ at a distance $r=d+R$. We obtain
as interaction energy:
\begin{align}
   V_{ps}&=-\hslash cB_{m,m}\frac{\pi}{10}\int_0^{2\pi}d\varphi
           \int_{-1}^1 d\cos\theta
           \int_0^R\dfrac{\rho^2d\rho}{(r+\rho\cos\theta)^4}\\
         &=-\hslash cB_{m,m}\frac{\pi^2}{30}
           \left(\frac{R}{d^2}-\frac{1}{d}+\frac{1}{d+2R}+
                 \frac{R}{(d+2R)^2}\right)
\label{310}
\end{align}
with:
\begin{equation}
   B_{m,m}\frac{\pi^2}{30}=0.04289
\label{320}
\end{equation}
This result should be compared with that of Balian and Duplantier
\cite{Duplantier} which in two scattering approximation for $d<<R$ is:
\begin{equation}
   V_{ps}\simeq -\frac{\hslash c}{8\pi}
         \left(\frac{R}{d^2}-\frac{1}{d}\right)
\label{330}
\end{equation}
where:
\[
    \left(\frac{1}{8\pi}=0.03978\right)
\]
We think that, apart from a few slight differences in the numerical 
coefficients, our expression can lead to a good approximation with 
respect to the exact calculation.

Although the dependence of the Casimir effect on the temperature is not
within the scope of the present brief note, we think appropriate to mention
at least the consequences that our Hamiltonian implies on the subject.
Our Hamiltonian includes diagonal terms which, for each body which may be 
present, generate a field energy equivalent to the body volume times the
black body radiation energy density as shown by Stefan-Boltzmann's
expression. But this is still a well known result. It is more interesting
to investigate on the effect of temperature on the dispersion forces. To
this end it is sufficient to insert as a factor of the integrand in equation  
\eqref{90} the quantities $\overline{n}(k',T)+\overline{n}(k,T)$
where $\overline{n}(k,T)$ is the average number of photons of wave
number $k$ at temperature $T$ obtained from Planck distribution. The
expression which comes out is not analytically menageable, however the 
insertion of the above mentioned factor introduces a natural cut-off which
implies that the quantity $U_{1,2}(T)$ can be easily evaluated using numerical
methods. However an analytical result can be obtained in the case of high
temperatures developing to the first order $\overline{n}(k,T)$ and
setting it equal to $KT/(\hslash ck)$. This procedure introduces 
a $k$ factor at the denominator of equation \eqref{90}. Then from
a power counting one can see that in this limit the interaction 
$U_{1,2}(T)$ between two volume elements at distance $r$ will behave 
as $1/r^6$. As a consequence in the case of two perfect metals 
in Casimir's configuration we will obtain, instead of expression  
\eqref{260}, an interaction energy proportional to $-KT(L^2/a^2)$ 
which is a well known result\cite{Mehra,Ballian}.

\end {document}